\documentclass[a4paper]{spie}  

\usepackage{txfonts}
\usepackage{amsfonts,amssymb}
\usepackage{graphicx}
\usepackage[colorlinks=true, allcolors=blue]{hyperref}
\usepackage{xspace}

\newcommand{\mic}{\ensuremath{\mu\mathrm{m}}\xspace}
\newcommand{\as}{\ensuremath{^{\prime\prime}}\xspace}
\newcommand{\lsd}{\ensuremath{\lambda/D}\xspace}

\title{On-sky compensation of non-common path aberrations with the ZELDA wavefront sensor in VLT/SPHERE}

\author[a]{A. Vigan}
\author[b]{M. N'Diaye}
\author[a]{K. Dohlen}
\author[c]{J. Milli}
\author[c]{Z. Wahhaj}
\author[d]{J.-F. Sauvage}
\author[a]{J.-L. Beuzit}
\author[a]{R. Pourcelot}
\author[e]{D. Mouillet}
\author[c]{G. Zins}

\affil[a]{Aix Marseille Univ, CNRS, CNES, LAM, Marseille, France}
\affil[b]{Universit\'e C\^ote d'Azur, Observatoire de la C\^ote d'Azur, CNRS, Laboratoire Lagrange, Bd de l'Observatoire, CS 34229, 06304, Nice Cedex 4, France}
\affil[c]{European Southern Observatory (ESO), Alonso de C\'ordova 3107, Vitacura, Casilla 19001, Santiago, Chile}
\affil[d]{Univ. Grenoble Alpes, CNRS, IPAG, 38000 Grenoble, France}
\affil[e]{ONERA, 29 Avenue de la Division Leclerc, 92322, Ch\^atillon Cedex, France}

\authorinfo{Send correspondence to A. Vigan: \href{mailto:arthur.vigan@lam.fr}{arthur.vigan@lam.fr}}

\begin{document}
\maketitle

\begin{abstract}
Circumstellar environments are now routinely observed by dedicated high-contrast imagers on large, ground-based observatories. These facilities combine extreme adaptive optics and coronagraphy to achieve unprecedented sensitivities for exoplanet detection and spectral characterization. However, non-common path aberrations (NCPA) in these coronagraphic systems represent a critical limitation for the detection of giant planets with a contrast lower than a few $10^{-6}$ at very small separations ($<$0.3\as) from their host star. In 2013 we proposed ZELDA, a Zernike wavefront sensor to measure these residual quasi-static phase aberrations and a prototype was installed in SPHERE, the exoplanet imager for the VLT. In 2016, we demonstrated the ability of our sensor to provide a nanometric calibration and compensation for these aberrations on an internal source in the instrument, resulting in a contrast gain of 10 at 0.2\as in coronagraphic images. However, initial on-sky tests in 2017 did not show a tangible gain in contrast when calibrating the NCPA internally and then applying the correction on sky. In this communication, we present recent on-sky measurements to demonstrate the potential of our sensor for the NCPA compensation during observations and quantify the contrast gain in coronagraphic data.
\end{abstract}



\section{Introduction}
\label{sec:introduction}

Circumstellar disks and planetary companions around nearby stars are routinely observed on the ground by several facilities with exoplanet direct-imaging capabilities \cite{Beuzit2008,Macintosh2008,Guyon2010c,Hinkley2011,Skemer2012,Close2014}. Of these facilities, the instruments VLT/SPHERE and Gemini Planet Imager (GPI) have seen first light in 2013-2014, providing unprecedented sensitivity and inner working angle for exoplanet observations \cite{Macintosh2014,vigan2015}. Since their commissioning, they have shed light on known or newly detected planetary companions with insights on their physical characteristics (orbit and mass) and atmospheric chemical features through spectral characterization and photometric and astrometric information \cite{galicher2014b,chilcote2015,vigan2016,zurlo2016}. In parallel, they are used to perform large surveys (GPIES, SHINE) that will target several hundreds of young, nearby stars, with the goal of probing the demography of the giant exoplanets population at large orbital separation. Up to now, these surveys have uncovered a few giant companions\cite{Macintosh2015,Konopacky2016,Chauvin2017,Ginski2018}, but many dozens of candidates remained to be confirmed and may lead to additional discoveries.

These ground-based instruments rely on a combination of extreme adaptive optics (ExAO) system, coronagraphy, dedicated observational strategies and post-processing methods. Differential aberrations between the ExAO sensing path and the science path, so-called non-common path aberrations (NCPA), have been identified as setting high-contrast performance limits for adaptive optics instruments. Their importance was well known \cite{Fusco2006} at the start of the development of the recently commissioned planet imagers, GPI and SPHERE, and various strategies were implemented to minimize them. For SPHERE, the very low order NCPA correction (tip, tilt and defocus) are optimised directly at the level of the coronagraph during the target acquisition, but the calibration strategy for higher orders, originally based on phase diversity techniques \cite{sauvage2007}, was not found to improve the final image quality and was finally discarded because the wavefront error budget was already within the specifications necessary to achieve the contrast objectives of the instrument. Still, the remaining NCPA are on the order of a few tens of nanometers, preventing coronagraphs from achieving their ultimate performance. These wavefront errors can be split into two contributions: the long-timescale aberrations that are due to the optical surface errors or misalignment in the instrument optical train and the slowly varying instrumental aberrations that are caused by thermal or opto-mechanical deformations as well as moving optics such as atmospheric dispersion correctors\cite{macintosh2005,martinez2012,Martinez2013}. They lead to static and quasi-static speckles in the coronagraphic images, which represent critical limitations for the detection and observation of older or lighter gaseous planets at smaller separations. More precise measurement strategies are required to measure and correct for these small errors with accuracy and achieve deeper contrast (down to $10^{-7}$, representing the ultimate contrast limit of these instruments) for the observation of the faintest companions.

For this purpose, we proposed the use of a Zernike phase mask sensor to calibrate the NCPA that are seen by the coronagraph in exoplanet direct imagers \cite{N'Diaye2013a}. A prototype of such a sensor, called ZELDA (Zernike sensor for Extremely Low-level Differential Aberrations), was finally implemented in SPHERE during the commissioning phase.  The first validation of ZELDA was presented in N'Diaye et al. (2016) and we demonstrated a clear potential to increase the raw contrast by a factor up to 10 at very small angular separation (0.1\as--0.4\as)\cite{N'Diaye2016}. However, these tests were done on the internal point source of the instrument during daytime, which provides an extremely stable environment for testing ultimate performance but is not necessarily representative of real observations. Indeed, the instrument being installed on a Nasmyth platform of the VLT-UT3, observations require the use of a derotator to stabilise either the field or the pupil. Also the fact that usual SPHERE/IRDIFS observations cover a broad range of wavelengths (from $Y$- to $K$-band) also requires the use of atmospheric dispersion correctors, which can also potentially introduce a small amount of NCPA. The next logical step was therefore to perform on-sky tests with the goal of measuring the gain in contrast provided by a proper compensation of the NPCA.

In this work, we present preliminary results of a series of on-sky tests performed with ZELDA on VLT/SPHERE. The final results will be provided in a forthcoming, more complete publication (Vigan et al. in prep.). In Sect.~\ref{sec:presentation_tests} we present a brief description of the strategy and technical aspects of the tests, then in Sect.~\ref{sec:loop_convergence} we compare the convergence of the NCPA compensation loop on the internal source and on sky, and in Sect.~\ref{sec:coronagraphic_perf} we present the coronagraphic performance with and without NCPA compensation on-sky. Finally in Sect.~\ref{sec:prospects} we present the conclusions and prospects of this work.

\section{Short presentation of the tests}
\label{sec:presentation_tests}

To perform on-sky tests with ZELDA, we benefited from 1.5 night of technical time awarded by ESO and the Paranal observatory. The tests were shared between daytime activities during 6 days and 3 half-nights on VLT-UT3 entirely dedicated to ZELDA validation. The goals for this test period were multiple: 

\begin{enumerate}
    \item checking previous results and performance that were obtained in 2015, \label{goal1}
    \item demonstrate that the ZELDA measurements can be performed on sky and used to compensate for NCPA, \label{goal2}
    \item measure the contrast gain in coronagraphic images in the presence of on-sky NCPA calibration, \label{goal3}
    \item compare internal and on-sky measurements, \label{goal4}
    \item define an operational strategy for a future implementation of NCPA calibration with ZELDA for all observations. \label{goal5}
\end{enumerate}

\noindent In the present work we focus on items \ref{goal2}, \ref{goal3} and \ref{goal4}, which are (to the best of our knowledge) completely new and have important implications for future instrumentation such as the high-contrast arm of ELT/HARMONI\cite{Thatte2016} or WFIRST/CGI\cite{Bailey2018}.

The ZELDA data analysis is entirely based on the public \texttt{pyZELDA} code\cite{Vigan2018a}. The approach for the tests follows a strategy very close to what was previously described in N'Diaye et al. (2016)\cite{N'Diaye2016} for the implementation of the ZELDA closed-loop operation, at the exception of the projection of the ZELDA optical path difference (OPD) map on the SPHERE deformable mirror (DM). As described in N'Diaye et al. (2016), the OPD map computed from ZELDA data cannot be directly projected on the DM because ZELDA measures spatial frequencies much higher than what the DM can actually correct for: up to 192 cycle/pupil (c/p) vs. up to 40 c/p, respectively. A direct projection of the OPD map on the DM will necessarily result in unpredictable results due to aliasing effects. In our previous work, we circumvented this issue by applying a low-pass spatial filtering on the OPD maps, with a cutoff frequency at 25 c/p, which was the value providing the best result at the time. Although effective, this approach was not the most elegant because it completely bypasses the Karhunen-Lo\`eve modes of the SPHERE AO system, SAXO\cite{Petit2014,Sauvage2016}. In the current approach, the OPD maps are first projected on the first 700 modes (out of 950+) that can be seen and controlled by SAXO, and finally being projected on the reference slopes of the WFS. This approach is hopefully more stable and less subject to noise than the previous one.

As presented in N'Diaye et al. (2016), another important part of the NCPA compensation with ZELDA is to make sure that the right amount of aberrations can be measured and then applied on the deformable mirror. This is what we called the \emph{sensitivity factor} of the wavefront sensor (WFS), which was previously calibrated using the introduction of a ramp of focus at calibrated amplitudes\cite{N'Diaye2016}. This calibration remains one of the basic pre-requirements of the NCPA calibration in SPHERE, but there are still several important questions that have not been addressed yet. In particular, the sensitivity factor displays a variability of up to 25\% depending at which time of the day the calibration is performed. This variability is not yet fully understood so for the time being the calibration is performed before all important test with ZELDA. Such an inaccuracy could be extremely problematic in open loop, because the applied correction would not correspond to the actual value. However, most of our tests are performed in closed loop, where this inaccuracy is acceptable, because the correction will always be performed in the proper direction as long as the amount of NCPA is small enough (see e.g. Vigan et al. 2011\cite{Vigan2011} for a similar scenario in the case of cophasing). For the moment we neglect the origin of the variation on the sensitivity factor, but keeping in mind that we would need to understand it for any open-loop use of the ZELDA sensor.

\section{Loop convergence}
\label{sec:loop_convergence}

Ideally, the NCPA are sufficiently small so that the ZELDA measurements are all within the quasi-linear range of the sensor, which enables to compensate for them in a single iteration. However, in practice the calibration issues described in Sect.~\ref{sec:presentation_tests} prevent to compensate for the NCPA in just 1 iteration, so that the compensation must be implemented in a closed-loop fashion that follows 5 distinct steps:

\begin{enumerate}
    \item obtain ZELDA measurement,
    \item compute the NCPA OPD map,
    \item filter the OPD map on the first 700 SAXO KL modes,
    \item project the filtered map on the WFS reference slopes,
    \item apply the new reference slopes on the WFS.
\end{enumerate}

\begin{figure}
  \centering
  \includegraphics[width=1.0\textwidth]{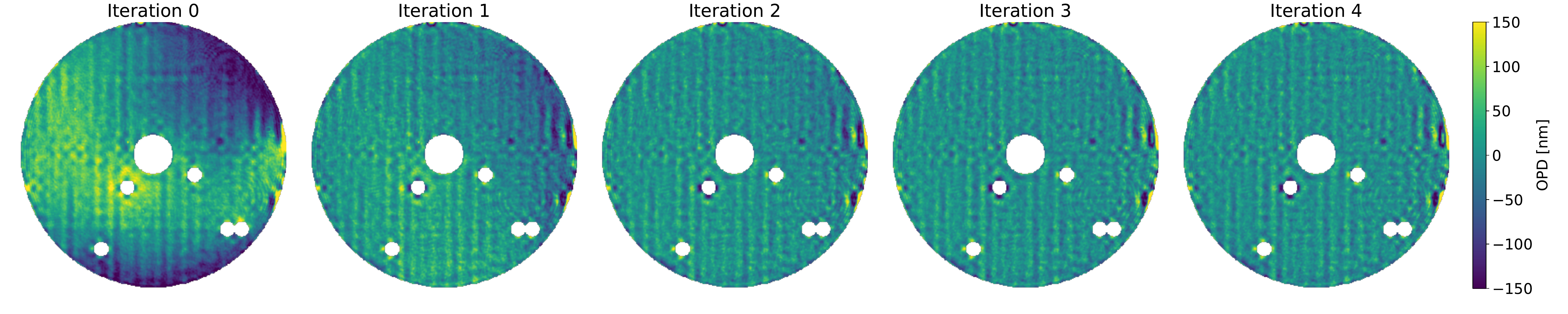}
  \includegraphics[width=0.5\textwidth]{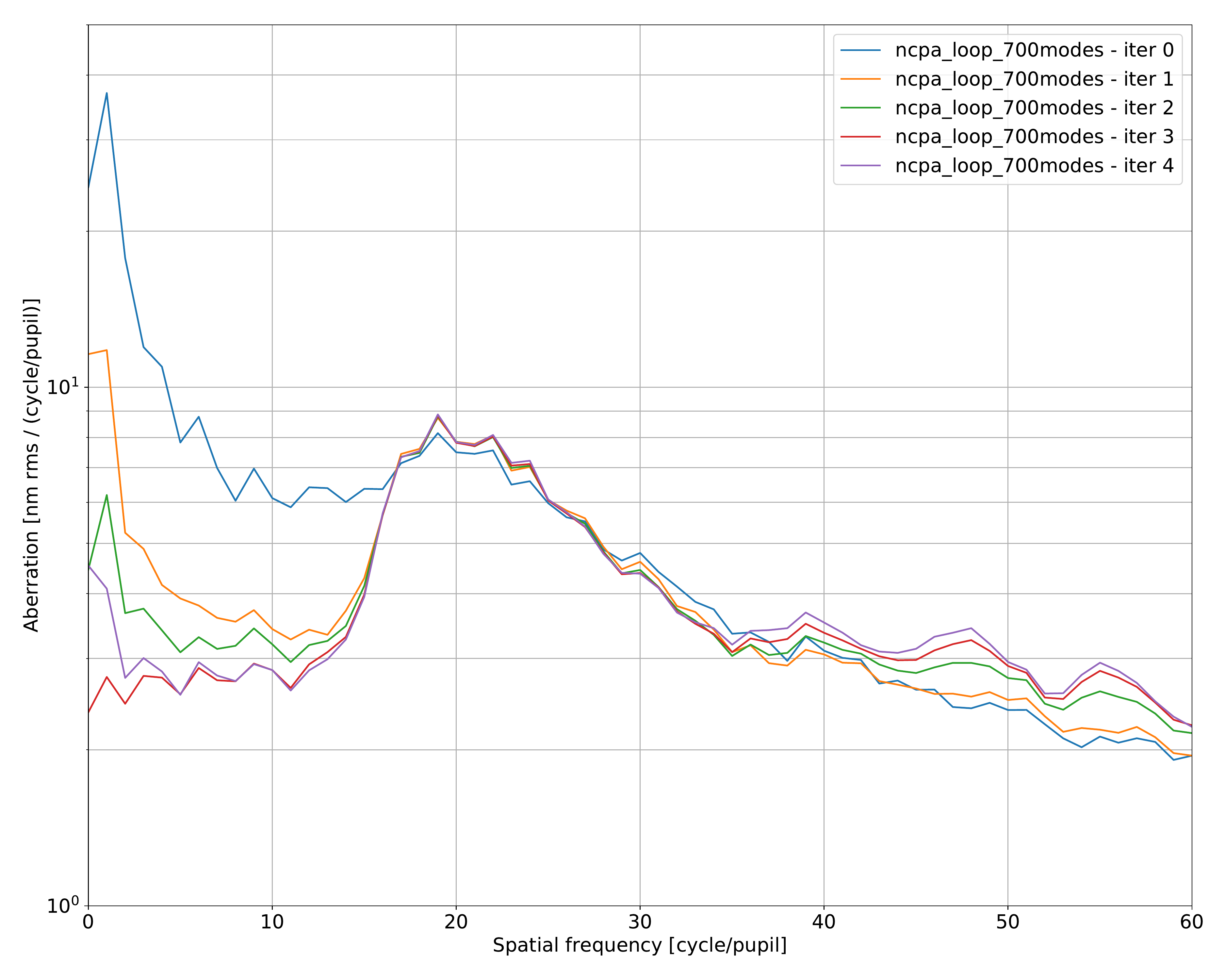}
  \caption{Convergence of the ZELDA loop on the internal source using 700 SAXO modes. \emph{Top}: OPD maps measured with ZELDA and calibrated in nanometers at all the iterations of the loop. The central part of the DM and the dead actuators have been masked in the OPD maps. \emph{Bottom}: Integrated power spectral density of the OPD maps as a function of spatial frequency for all the iterations. The aberration value provided at spatial frequency $s$ is equal to the value of the PSD integrated between $s$ and $s+1$ cycle/pupil. The central part of the DM and the dead actuators have also been masked in this analysis.}
  \label{fig:int_convergence}
\end{figure}

Figure~\ref{fig:int_convergence} shows the convergence of the ZELDA loop on the internal source of VLT/SPHERE. The top figure displays the OPD maps computed at the 5 iterations of the loop, while the bottom plot displays the integrated power spectral density (PSD) computed on each of the OPD maps. The plot is not directly the raw PSD, but the PSD integrated within bounds of width 1 c/p: the aberration value provided at spatial frequency $s$ is equal to the value of the PSD integrated between $s$ and $s+1$ cycle/pupil. This enables to directly obtain an estimation of the aberrations, in nm rms, in a given spatial frequency range. The total amount of NCPA can be obtained by computing the quadratic sum of all values. In this analysis, the central part of the DM, where the actuators are not controlled, and the dead actuators of the mirror have been masked.

The evolution of the OPD maps in Fig.~\ref{fig:int_convergence} clearly shows a visual improvement of the NCPA: the strong astigmatism that clearly dominates at start quickly disappears to only leave a relatively flat wavefront dominated by the print-through of the actuators and some high-spatial frequencies at the right edge of the DM where several dead actuators prevent from a completely clean correction. More quantitatively, the integrated PSD shows a decrease by factor of more than 10 in the 1-2~c/p bin after 3 iterations, and a decrease by a factor 2-3 in the 4-12~c/p bins. There is no visible effect for spatial frequencies above 15~c/p because of the filtering of the OPD maps on the first 700 SAXO KL modes. Between iteration 0 and iteration 3, the amount of aberrations goes from 53~nm rms down to 12~nm~rms in the 1-15~c/p range, and from 48~nm rms down to 6~nm rms in the 1-4~c/p range.

\begin{figure}
  \centering
  \includegraphics[width=1.0\textwidth]{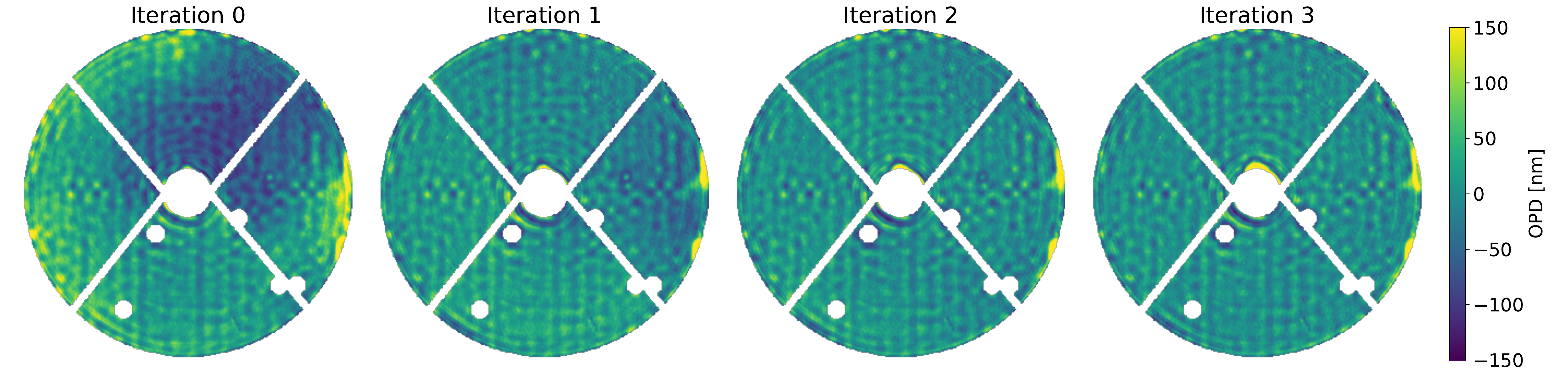}
  \includegraphics[width=0.5\textwidth]{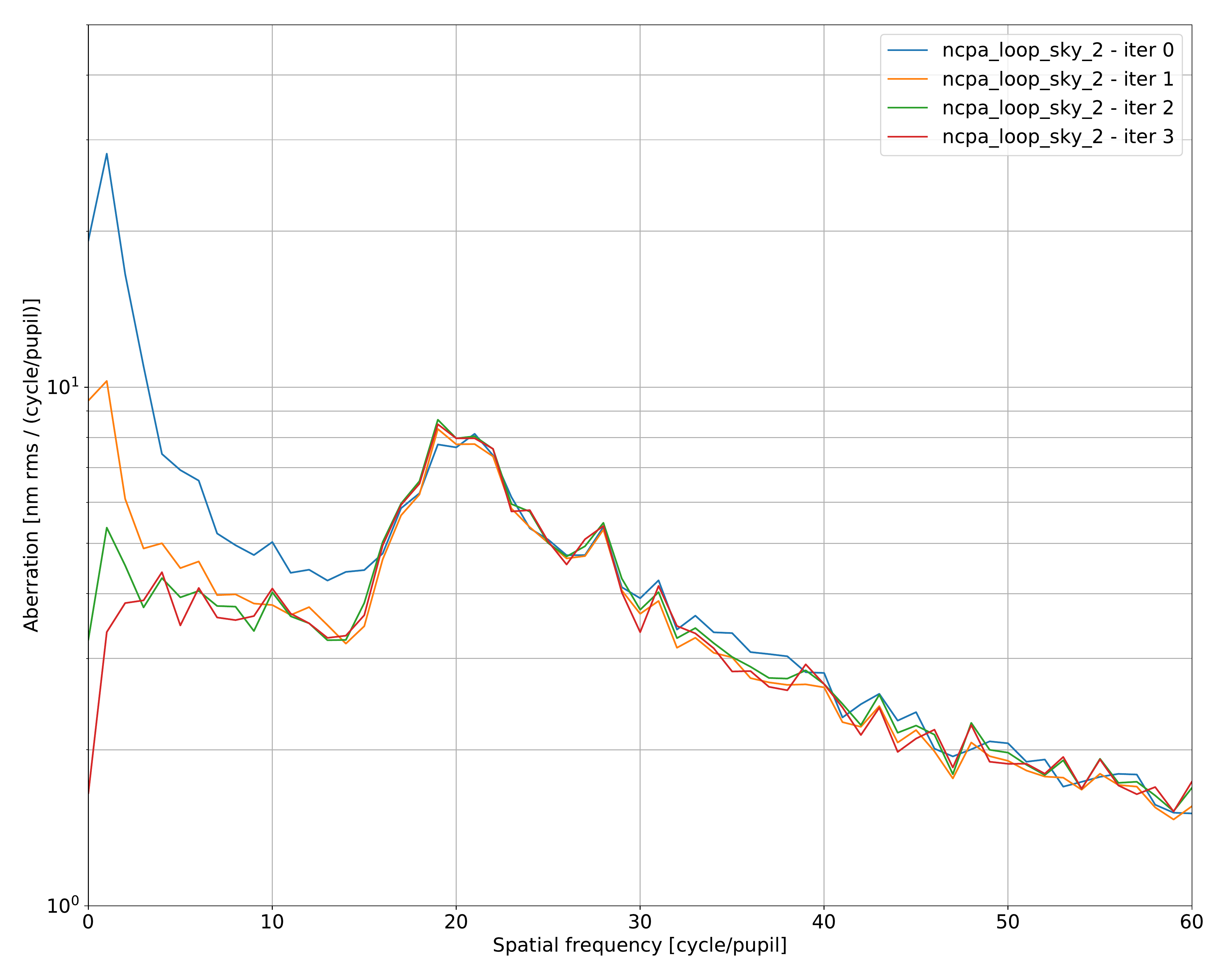}
  \caption{Same as Fig.~\ref{fig:int_convergence} but the data is now acquired on-sky on a bright star ($\alpha$ Crt, K0, $V=4.07$, $H=1.76$) observed on 2018-04-03. In these on-sky data, the spiders have been masked in the analysis in additional to the central obscuration and the dead actuators of the DM.}
  \label{fig:sky_convergence}
\end{figure}

Figure~\ref{fig:sky_convergence} shows the same results as Figure~\ref{fig:int_convergence} but this time on-sky on star $\alpha$ Crt, a very bright K0 star ($V=4.07$, $H=1.76$) which was observed at high elevation in relatively poor observing conditions on 2018-04-01 (1.0\as-1.2\as seeing, coherence time $<$3~ms). The OPD maps look relatively similar to the maps on the internal source, except for the presence of the spiders holding the secondary mirror. Some mid-spatial frequency circular phase structures from the primary mirror are also clearly visible. They correspond to polishing errors on the telescope primary mirror due to the original machining. The integrated OPD curve shows a very similar evolution as on the internal source, although the starting point in the 5-15~c/p range seems lower for these on-sky measurements. The reason for this different starting point is not completely understood yet. In these measurements, between iteration 0 and iteration 3, the quantity of aberrations goes from 39~nm rms down to 14~nm~rms in the 1-15~c/p range, and from 35~nm rms down to 8~nm rms in the 1-4~c/p range.

These initial on-sky results are extremely encouraging and demonstrate that NCPA measurement and compensation in presence of ExAO-filtered atmospheric residuals are possible. The differences with the internal calibration will be investigated in the coming months, for example to understand whether or not the difference on the starting point is real or the result of a measurement bias on sky. An interesting aspect of these tests is also that it does not seem necessary to average the atmospheric residuals for integration times longer than 15-20~sec. Indeed, we perform additional tests (not present here) which did not show any visible differences in the PSD of the measured NCPA with integration times from 15~sec up to 120~sec. This means that for bright stars, for which we would expect the NCPA correction to bring a visible improvement on the quasi-static speckles, a short overhead of only $\sim$1~min would be necessary at the beginning of the observation sequence.

\section{Coronagraphic performance}
\label{sec:coronagraphic_perf}

The validation of the NCPA correction was performed on 2018-04-01 by switching to coronagraphic imaging immediately after the NCPA calibration sequence. For the coronagraph, We used an APLC\cite{Soummer2005} with a focal plane mask of 185~mas in diameter. The images were acquired in the IRDIS/DBI mode with the $H2$ filter at $\lambda = 1.593$~\mic in relatively poor observing conditions with \texttt{NEXP}$\times$\texttt{DIT}$\times$\texttt{NDIT}=4$\times$7$\times$10~sec. Unfortunately the 10 images for each of the 4 exposures were averaged at the level of the detector controller, which means that we do not have access to the individual exposures. Since the observing conditions were far from optimal, it is possible that some images with poor AO correction were averaged in the sequence, making it difficult to estimate the absolute gain provided by the NCPA correction.

\begin{figure}
  \centering
  \includegraphics[width=0.8\textwidth]{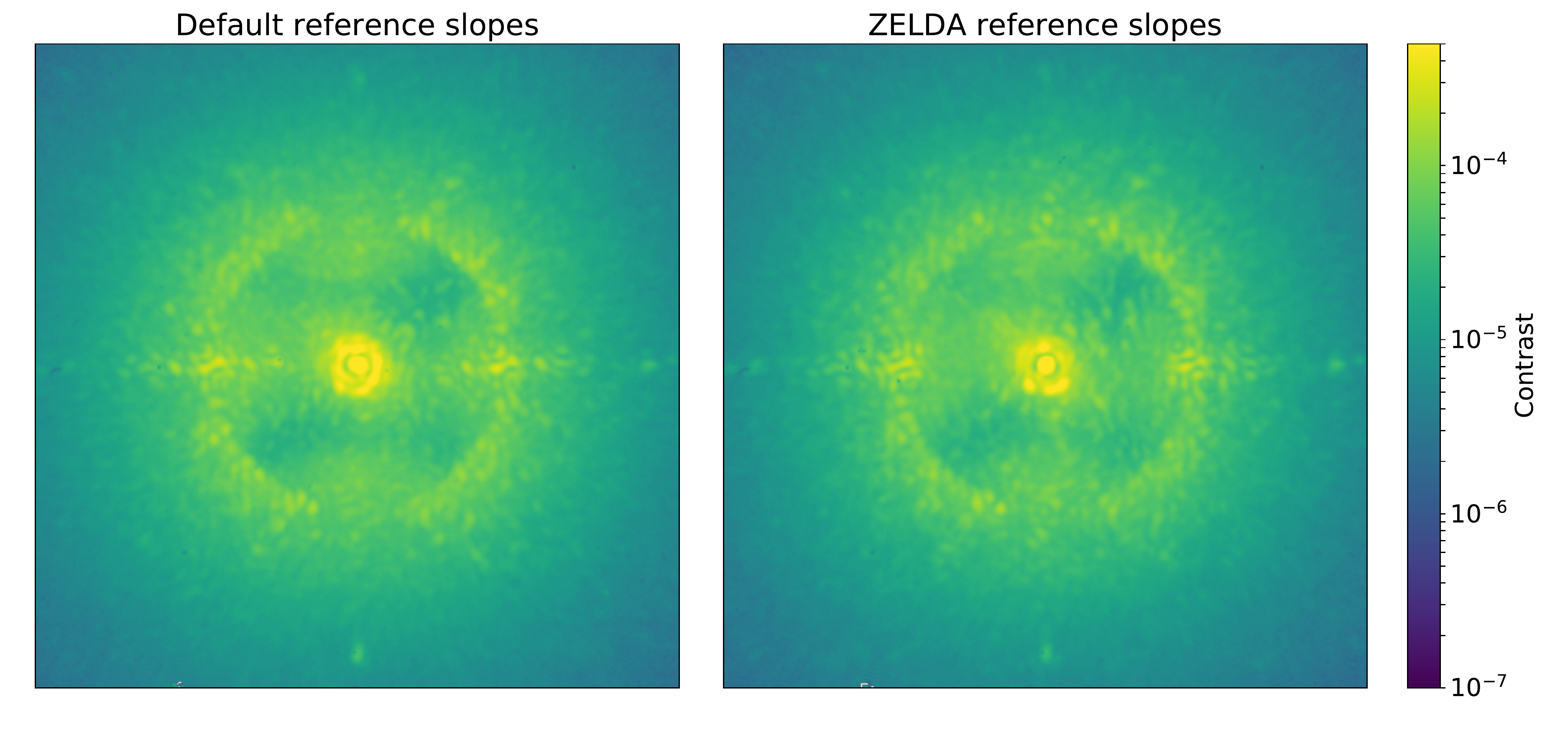}
  \includegraphics[width=0.7\textwidth]{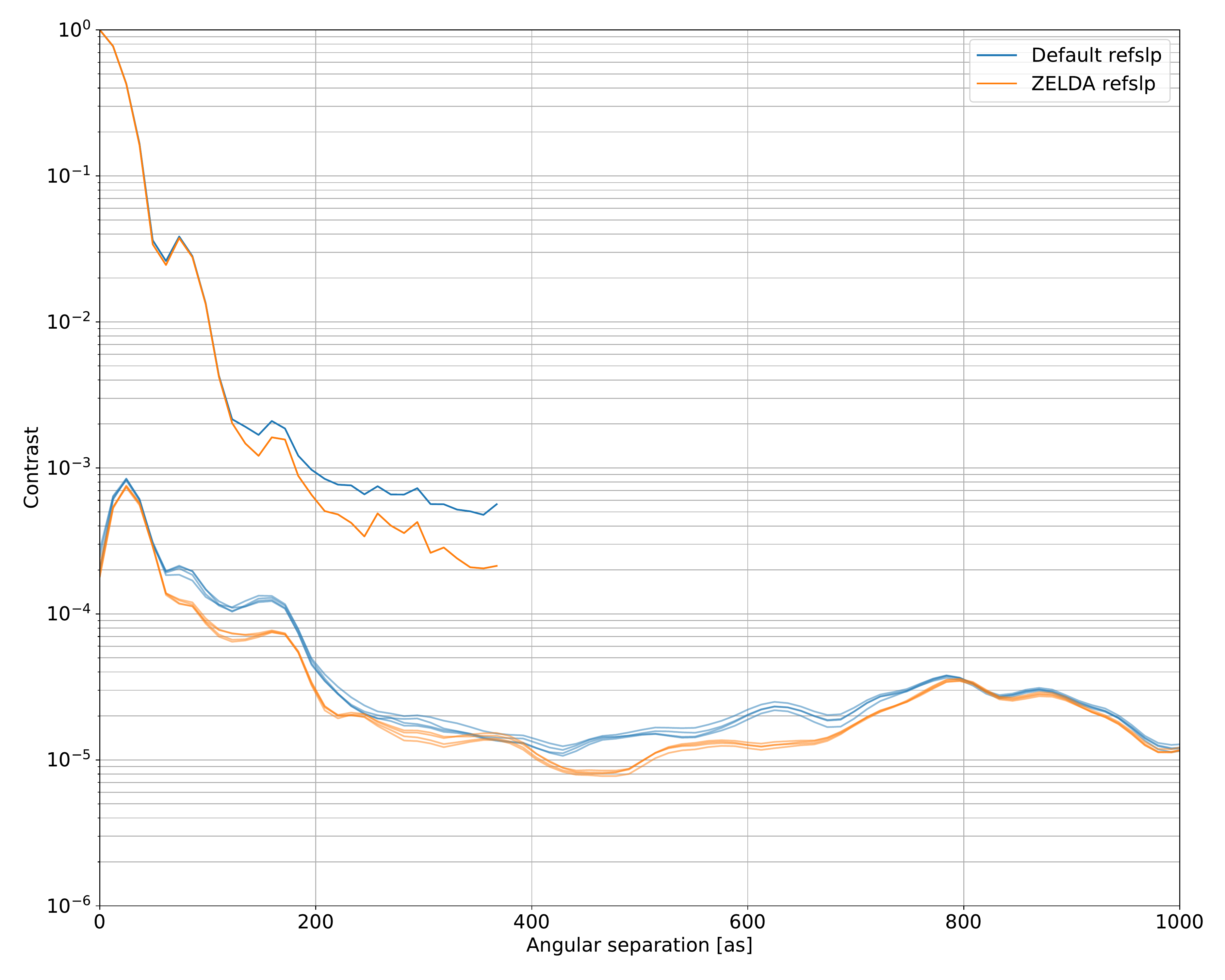}
  \caption{Coronagraphic performance on sky based on a sequence acquired on 2018-04-01 on $\alpha$ Crt (K0, $V=4.07$, $H=1.76$). \emph{Top}: coronagraphic images, calibrated in contrast with respect to the peak of unocculted PSF, with the default WFS reference slopes and with the reference slopes corrected to compensate the NCPA based on the ZELDA measurement after 3 iterations (see Fig.~\ref{fig:sky_convergence}). The coronagraph is an APLC with a focal plane mask of 185~mas in diameter. The images are acquired in the IRDIS/DBI mode with the $H2$ filter at $\lambda = 1.593$~\mic. The data were acquired in relatively poor observing conditions (1.0\as-1.2\as seeing) with \texttt{NEXP}$\times$\texttt{DIT}$\times$\texttt{NDIT}=4$\times$7$\times$10~sec. Unfortunately the 10 images for each of the 4 exposures were averaged at the level of the detector controller, which means that we do not have access to the individual exposures and we cannot select the best frames for the contrast estimation in these poor observing conditions. \emph{Bottom}: Contrast curves corresponding to the two images. The contrast is calculated as the azimuthal standard deviation in annulii of width \lsd.
  }
  \label{fig:sky_coro}
\end{figure}

The results are presented in Fig.~\ref{fig:sky_coro}. The top row of the figure shows coronagraphic images without NCPA compensation (left, iteration 0 in Fig.~\ref{fig:sky_convergence}) and with NCPA compensation (right, iteration 3 in Fig.~\ref{fig:sky_convergence}). These images are normalized to the peak flux of an off-axis reference PSF of the star acquired at the beginning of the sequence. The bottom row of the figure show the contrast curves for the images acquired at each loop iteration, again normalized to the intensity peak of the off-axis stellar PSF. The contrast curves are calculated as the azimuthal standard deviation in annulii of width \lsd. 

The gain in contrast is not obvious by just looking at the coronagraphic images. Speckles at the edge of the coronagraph are clearly modified by the compensation of the NCPA, but it is hard to tell whether their variance is decreased or not. The horizontal and vertical radial structures caused by the print-through of the actuators of the DM are clearly attenuated up to separations of $\sim$15\lsd. The contrast plot shows a small gain in contrast between 100 and 200 mas and between 400 and 700 mas, but this gain remains below a factor of 2, which is marginal.

\begin{figure}
  \centering
  \includegraphics[width=0.7\textwidth]{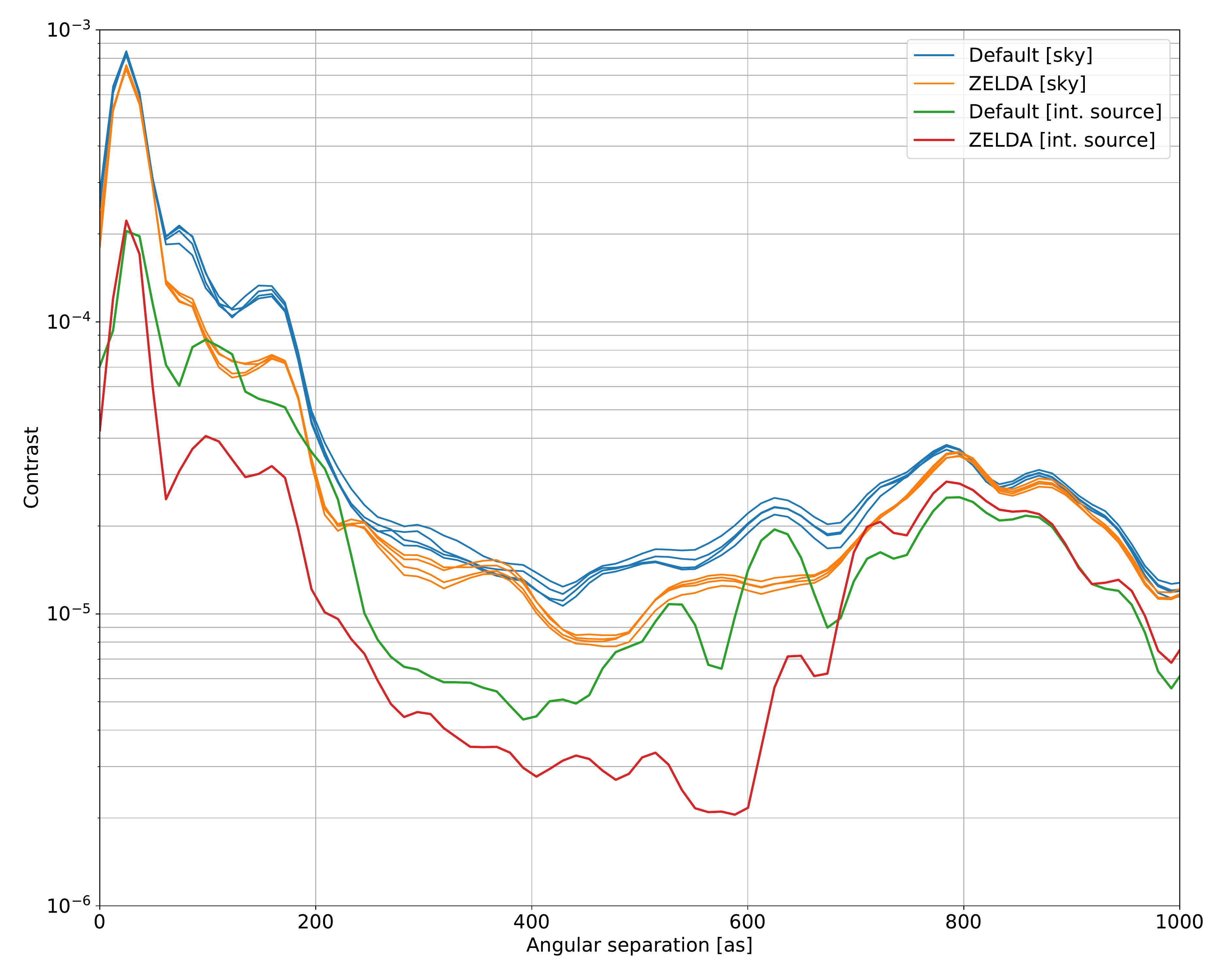}
  \caption{Comparison of the contrast curves with and without compensation of the NCPA on the internal source and on sky based on the data acquired on 2018-04-01. The on-sky data is compared to internal data acquired during the afternoon preceding the observations. For the on-sky data, the 4 curves correspond to the contrast obtained on the 4 different exposures acquired on 2018-04-01. The are the same as the curves presented in Fig.~\ref{fig:sky_coro}}
  \label{fig:int_sky_coro}
\end{figure}

The comparison between the gain on the internal source and on sky is presented in Fig.~\ref{fig:int_sky_coro}. However, the on-sky data is most likely limited by the uncorrected atmospheric residuals. This is expected, especially in poor observing conditions. Temporal residuals close to the axis seem to be much higher on-sky, as well as the ``speckle floor'' between 200 and 600~mas. Nonetheless some gain is visible at similar separations both on-sky and internally, which confirms that the NCPA compensation really has an effect on the quasi-static speckles, but because of the poor conditions it is impossible to exactly estimate the final contrast level set by these speckles. The data remain dominated by uncorrected AO-residuals.

\section{Conclusions \& prospects}
\label{sec:prospects}

High-contrast imaging of exoplanets from the ground requires to combine extreme AO and coronagraphy, but for optimal coronagraphic attenuation one needs to properly measure and compensate for the NCPA between the wavefront sensing and the science paths in the instrument. The ZELDA wavefront sensor implemented in VLT/SPHERE enables to calibrate some with nanometric accuracy. 

The performance of this sensor for NCPA compensation has been previously demonstrated. In our new results, the quantitative estimation of the PSD of the OPD maps measured with ZELDA demonstrates that the low-spatial frequency aberrations (1-4 c/p) can be reduced by a factor 8 on the internal source, which translates into a gain of more than a factor 3 in contrast at 200 mas, and almost 10 at 600 mas. These results are slightly worse than the ones that were previously obtained in 2015\cite{N'Diaye2016}. This could potentially be explained by the fact that in 2015 the defocus at the level of the coronagraph mask may not have been compensated in the data without NCPA compensation, resulting in worse coronagraphic performance in our recent data. In the new data, particular care was taken to ensure an optimal performance in every configuration for fair comparison.

We also present the very first on-sky NCPA compensation using ZELDA. The analysis of the OPD maps and the corresponding PSD shows a clear gain, at an equivalent level to the one obtained on the internal source. The final aberration level seems to be almost identical between the internal source and on-sky. On the coronagraphic images, the gain is less obvious. Some speckles are clearly affected by the NCPA compensation, but the gain in contrast remains marginal. However, a gain in contrast is observed at similar separations both internally and on-sky, which is a good indication that the NCPA are indeed properly compensated. The fact that the observing conditions were relatively poor when the data were acquired almost certainly explains the small gain in contrast: the data remain dominated by uncorrected AO-residuals. From this we conclude that under poor to fair seeing and coherence time conditions, the calibration of the NCPA brings only a marginal gain as quasi-static speckles do not constitute the main limitation in the well-corrected region of the focal plane. Therefore, use of night-time for the calibration of the NCPA would only make sense in moderate to good corrections. Such a scheme would fit well within the new set of atmospheric constraints including coherence time that will be implemented for service mode observing on VLT/SPHERE in 2019, which will give the possibility for astronomers to request excellent to good conditions based on coherence time.

Additional ZELDA data and coronagraphic data have been acquired during our tests at the ESO/Paranal observatory, which hopefully will provide more insight into the current limitations of ZELDA on-sky. In particular some data will provide information regarding the stability of the NCPA compensation in real observing conditions, where the telescope is moving and the derotator and ADCs are rotating. This temporal aspect is essential to define operational aspects like the frequency at which the NCPA calibration must be executed to bring a real benefit to the observations. All these results will be presented in a forthcoming publication.

\acknowledgments

This project has received funding from the European Research Council (ERC) under the European Union's Horizon 2020 research and innovation programme (grant agreement No. 757561). AV and MN would like to thank ESO and the Paranal observatory for their strong support during their visitor run and their continued support for the implementation of ZELDA as the NCPA calibration strategy for SPHERE.

\smallskip

SPHERE is an instrument designed and built by a consortium consisting of IPAG (Grenoble, France), MPIA (Heidelberg, Germany), LAM (Marseille, France), LESIA (Paris, France), Laboratoire Lagrange (Nice, France), INAF - Osservatorio di Padova (Italy), Observatoire de Genève (Switzerland), ETH Zurich (Switzerland), NOVA (Netherlands), ONERA (France) and ASTRON (Netherlands) in collaboration with ESO. SPHERE was funded by ESO, with additional contributions from CNRS (France), MPIA (Germany), INAF (Italy), FINES (Switzerland) and NOVA (Netherlands). SPHERE also received funding from the European Commission Sixth and Seventh Framework Programmes as part of the Optical Infrared Coordination Network for Astronomy (OPTICON) under grant number RII3-Ct-2004-001566 for FP6 (2004-2008), grant number 226604 for FP7 (2009-2012) and grant number 312430 for FP7 (2013-2016).

\bibliography{paper}
\bibliographystyle{spiebib_short}

\end{document}